\RequirePackage{fix-cm}
\documentclass{svjour3}
\smartqed

\usepackage{graphicx}
\usepackage{color}
\usepackage{url}
\usepackage{amsfonts}
\usepackage{dirtytalk}
\usepackage{amsmath}
\usepackage{bbm}

\newcommand{\new}[1]{{\color{black}{#1}}}

\newcommand{\anony}[1]{#1}

\begin{document}

\title{What Should We Optimize in Participatory Budgeting? An Experimental Study}

\anony{
\author{Ariel Rosenfeld \and Nimrod Talmon}
\institute{Ariel Rosenfeld \at
              Bar-Ilan University, Israel \\
              \email{ariel.rosenfeld@biu.ac.il}           
           \and
           Nimrod Talmon \at
              Ben-Gurion University, Israel\\
              \email{talmonn@bgu.ac.il}
}
}

\date{Received: date / Accepted: date}

\maketitle

\begin{abstract}
Participatory Budgeting (PB) is a process in which voters decide how to allocate a common budget; most commonly it is done by ordinary people --  in particular, residents of some municipality -- to decide on a fraction of the municipal budget. From a social choice perspective, existing research on PB focuses almost exclusively on designing computationally-efficient aggregation methods that satisfy certain axiomatic properties deemed \say{desirable} by the research community.
        Our work complements this line of research through a user study ($N = 215$) involving several experiments aimed at identifying what potential voters (i.e., non-experts) deem fair or desirable in simple PB settings. Our results show that some modern PB aggregation techniques greatly differ from users' expectations, while other, more standard approaches, provide more aligned results. We also identify a few possible discrepancies between what non-experts consider \say{desirable} and how they perceive the notion of \say{fairness} in the PB context. Taken jointly, our results can be used to help the research community identify appropriate PB aggregation methods to use in practice.
\end{abstract}

\section{Introduction}\label{section:introduction}

Participatory Budgeting (PB)~\cite{participatoryBudgeting} is a process in which voters decide how to allocate a common budget. It was first implemented in 1989, by the Brazilian municipality of Porto Alegre. Since then, it had widely spread to more than 1,500 municipalities worldwide, including major European capitals such as Madrid, Paris, Berlin, and Warsaw, as well as to other municipalities in Latin and North America, Asia, Australia, South Africa, and the Middle-east~\cite{participatoryBudgeting,shah2007participatory,PBO}.  PB  usually operates as follows:
  first, the city council declares a portion of its annual budget to be assigned to the PB process.
  Then, city residents can suggest projects, such as, e.g., bicycle routes, certain renovations, playgrounds, etc.
  The city then filters the list of all proposals by the residents and assigns a price tag to each of them.
  Then, in the second phase of the process, residents act as voters by specifying their preferences over the projects, usually by submitting approval ballots (e.g., each resident can choose at most $5$ projects to vote for). Then,  an aggregation method (i.e., a voting rule) is used by the city to select a subset of the proposed projects to fund.
  
From a social choice perspective, PB is usually formulated as follows:
  given a set of projects $P$ with their associated cost function $C : P \to \mathbb{N}$, a set of voters~$V$ with their associated approval ballots, and a budget limit $\ell$; the aggregation task is to output a \emph{bundle} $B \subseteq P$ satisfying $\sum_{p \in B} C(p) \leq \ell$.
For this model, a wide variety of mathematical formulations for PB were proposed, investigated, and addressed through either an axiomatic and/or an algorithmic approach (see \cite{aziz2020participatory} for a recent survey). Specifically, existing research on PB focuses almost exclusively on designing computationally-efficient aggregation methods that satisfy certain axiomatic properties deemed \say{fair} or \say{desirable} by the research community.
For example, Fain et al.~\cite{FainGM16} aim at satisfying a proportionality axiom related to the game-theoretic concept of the \emph{core} (see also their generalization~\cite{fainMunSha18:core_and_pb});
Flushnik et al.~\cite{fairknapsack} and Peters et al.~\cite{peters2020proportional} consider PB when voters provide cardinal utilities;
Aziz et al.~\cite{aziz2018proportionally} propose an aggregation method for PB with approval ballots that satisfies a proportionality axiom related to the axiom of Justified Representation for multiwinner elections~\cite{mwchapter} (see also the related paper regarding ordinal ballots~\cite{aziz2019proportionally});
Benade et al.~\cite{benade2017preference} choose among different ballot types for PB, based on a notion of \emph{distortion} (which, roughly speaking, asks how \say{close} an aggregation based on each ballot type is to the best outcome that could be computed if the true utilities were known);
Faliszewski and Talmon~\cite{abpb} study various approval-based aggregation methods for PB, concentrating on certain monotonicity axioms;
Jain et al.~\cite{jain2020participatory} generalize the model of Faliszewski and Talmon by considering interactions between projects.

\new{To our knowledge, most of the existing PB literature has yet to consider a user study perspective, focusing almost exclusively on axiomatic and/or algorithmic approaches to PB (indeed, we mention some related work that does so below)}.
Our work thus complements the existing PB research through a user study, involving two experimental settings (overall, $N = 215$), aimed at identifying what potential voters (i.e., non-experts) deem desirable in simple PB settings.
In particular, we seek to offer answers to the following two questions:
\begin{enumerate}

\item
Assuming that voters get some sort of \emph{utility} from each project, and assuming that these values are known, then what function of these utilities should be optimized when choosing a bundle of projects to fund, according to the views of common people?

\item
Assuming that voters only provide approval ballots, and assuming that such approval ballots represent some underlying per-project utilities, how shall these \emph{utilities} be estimated, according to the views of common people?

\end{enumerate}

\new{Indeed, our approach in this study is a utility-based one, following the general utility-based approach commonly used in game theory and decision theory.
In the context of PB, we mention the paper of Talmon and Faliszewski~\cite{abpb}, which study several aggregation methods for PB through the definition of several \emph{satisfaction functions}; these satisfaction functions essentially translate a voter ballot together with a possible output bundle into a numerical value corresponding to the expected \emph{satisfaction} of the voter for the bundle, which is closely related to our second experiment. In this context, in our first experiment we assume that we are given these per-project satisfaction values explicitly, and some of these are adopted in our second experiment as well.}
 
To this end, we report on set of experiments considering two different settings: First, settings in which the utilities each voter receives from each project are objective, additive, and known; given such utilities, we ask non-experts to choose the most appropriate bundle to fund. Second, settings in which only the approval ballots of voters are known, and we ask non-experts to choose the most appropriate bundle to fund.\footnote{All experiments were authorized by the corresponding IRB.}
We use the results to identify which aggregation methods common people view as the most appropriate in known-utilities cases and in approval-ballots cases.

Our results show that recently proposed aggregation methods, whose development is in the focus of the research effort on PB, need not necessarily align with what non-experts consider appropriate or desirable. Instead, we identify a few, quite standard approaches that appear to be more adequate. We further identify a few possible discrepancies between what non-experts consider \say{desirable} and how they perceive the notion of \say{fairness}, between men and women and between people of different economical statuses.  
Our results can be instrumental in the selection of appropriate PB aggregation methods in practice and highlight various promising avenues for further research for the community. 

There are 

This study is inspired by a variety of user studies performed in the larger context of computational social choice research, which have helped to better situate and understand the theoretical and computation advances made in various settings. 
For example, human voting behavior has been the focus of various experimental studies such as that by
Scheuerman~\cite{scheuerman2019heuristic,scheuerman2020modeling}, who has identified different manipulation aspects in human approval voting; 
Tal et al.~\cite{tal2015study}, who have demonstrated different voting behaviors in various online settings under the Plurality rule;
Zou et al.~\cite{zou2015strategic}, who have studied voter behavior in Doodle pools; 
and
Grandi et al.~\cite{grandi2020voting}, who have studied iterative human voting in combinatorial domains.
\new{In a very broad context, some studies in experimental economics share a common goal in quantifying justice or fairness (see, e.g.,~\cite{konow2016economics}).
In the context of fair division, we mention the work of Herreiner and Puppe~\cite{herreiner2009envy} as well as that of Gal et al.~\cite{gal2016fairest}.}
Furthermore, in the realm of matching/assignment problems, user studies have helped in understanding non-truthful behaviors in corresponding non-truthful mechanisms (e.g., \cite{artemov2017strategic}), \say{almost} truthful mechanisms (e.g., \cite{rosenfeld2020too}),  and truthful mechanisms (e.g., \cite{rees2018experimental}). 
Common to these and similar works is the understanding  that any intelligent automated decision, let alone one that involves potentially millions of dollars, necessitates the understanding of real-world users (in our case, voters/residents) and their judgment (in our case, what people deem appropriate) \cite{rosenfeld2018predicting}. \new{Closely related to our work is the recent study by Bulteau et al.~\cite{9475498}, who have studied theoretically, empirically and through a human study, various notions of justified representation in perpetual voting settings. }
In the context of PB, we mention several works that deal with user experiments:
Goel et al.~\cite{goel2019knapsack} study behavior of voters using Knapsack ballots for PB;
Laruelle~\cite{laruelle2020voting} studies a specific PB instance;
and Benade et al.~\cite{benade2018efficiency} conducted a user experiment to compare different ballot types for PB.



\section{Preliminaries}

In this paper, we consider the standard model of PB, also referred to as \emph{combinatorial PB} \cite{aziz2020participatory}. We assume a set $P = \{p_1, \ldots, p_m\}$ of projects, with an associated cost function $C : P \to \mathbb{N}$, such that $C(p)$ is the cost of project $p \in P$. For simplicity, we slightly overload notation by defining $C(P') := \sum_{p \in P'}C(p)$ where $P'\subseteq P$.
Furthermore, there is a set $V = \{v_1, \ldots, v_n\}$ of voters, where each voter $v_i$ reports her preferences over the projects. Again, we overload notation and refer to the preference of voter~$v_i$ as $v_i$ as well.

In this study we focus on two common preference types:
\begin{itemize}

\item
\textbf{Cardinal utilities:}
here, $v_i$ is a utility function, such that $v_i(p)$ is the utility gained by voter $i$ if project $p$ were to be funded. As an example, one can consider the expected revenue for a store if a certain project were to be funded.
We assume additive utilities (in particular, we assume that no project interactions, like those assumed by Jain et al.~\cite{jain2020participatory} are present).

\item
\textbf{Approval ballots:}
here, each voter specifies a subset of the projects that she \say{accepts}; that is, $v_i \subseteq P$, so that $p \in v_i$ if $v_i$ \emph{approves} $p$ and $p \notin v_i$ if $v_i$ \emph{disapproves} $p$.

\end{itemize}

Given a budget limit $\ell$, an \emph{aggregation} method is given a tuple $(P, C, V, \ell)$ and is required to output a \emph{winning bundle} $B \subseteq P$, with $\sum_{p \in B} c(P) \leq \ell$, containing the projects that should be funded.\footnote{Ties are broken  arbitrarily.}



\section{Experiment 1:  PB with Cardinal Utilities}\label{section:experiment1}

In this experiment, we examine the case of PB with real-valued utilities. (In a way, this experiment relates to general aggregation of utilities; we chose to conduct this experiment as the setting of PB is specific and thus may have different properties; furthermore, we are not aware of experiments in the general setting that are similar enough to the aim of this experiment.) Specifically, we devise a simple yet natural PB scenario where the utilities each voter receives from each project are objective, additive, and known; given such utilities, non-experts are tasked with choosing the most appropriate bundle to fund. Specifically, we use the following motivational scenario in our experiment:

\smallskip\noindent\textit{\textbf{Motivating Scenario (Cardinal Utilities):}
Alice is the manager of a small shopping mall containing 5 stores.
Alice just got a municipal grant of NIS 10M and wishes to use it to increase the revenues of the different stores in it.
Alice cannot invest more then NIS 10M and unused money will go back to the city administration.
Alice was offered 5 possible projects (enlarging the parking space, adding an elevator, etc.), where each project has a different cost and will provide different revenue boosts of each project to each of the stores. An impartial  economic advisor has estimated the costs of each project as well as the expected revenue boost to each of the stores.
Alice's job is to choose which projects to fund.\footnote{The \say{constants} of the scenario (budget of NIS 10M and 5 stores in the mall) were chosen empirically following a short informal trial-and-error investigation with students in our labs.}
}
\smallskip

\begin{figure}[ht]
\centering
\includegraphics[width=0.85\textwidth]{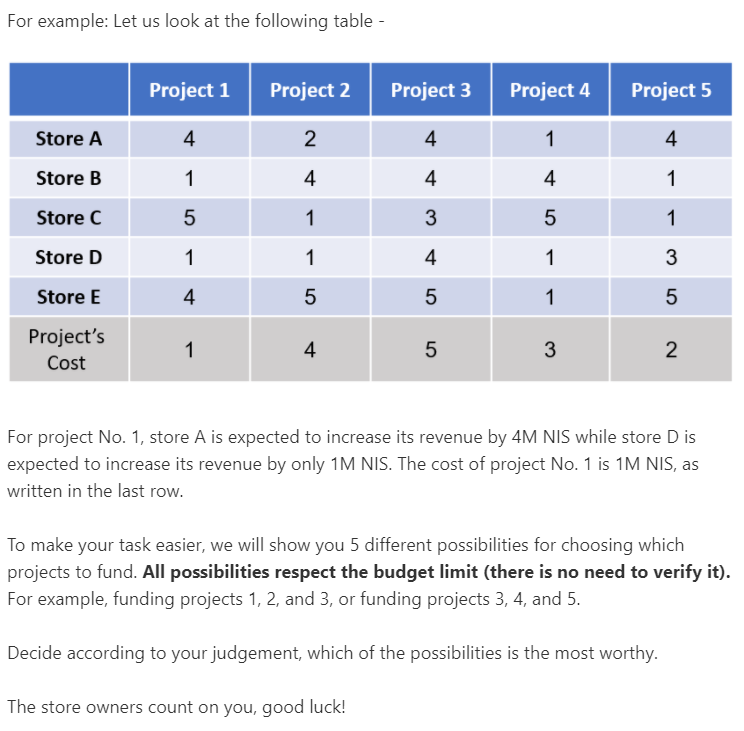}
\caption{Experiment 1: An example instance presented to our participants.}
\label{fig:SS}
\end{figure}

The above scenario was devised following this rationale: by using monetary revenues for stores, which are assessed by an independent third party, we can fairly assume that potential decision makers would consider the utilities as being objective, additive, and known. In addition, since Alice's utility is not defined at all (she is only introduced as the mall's manager), we leave it up to the decision maker to decide what she constitutes as appropriate in this generic setting, without introducing potentially biasing terms such as \say{fairness}. 

Following the motivating scenario's text, participants were presented with an example instance, as shown in Figure \ref{fig:SS}.


\subsection{Experiment 1: Design}

Following the motivational scenario outlined above, we first randomly generated a set of instances, each including voter preferences and projects' costs, in which the projects' costs were randomly drawn from a uniform distribution over the set \{1,2,3,4,5\}. 
To examine different underlining aggregation axioms and principles, each generated instance was first solved using five different aggregation methods, as detailed next. Then, five instances for which none of the five output bundles coincide were selected (the ordering by which these instances were shown to our participants was random to each participant). Specifically, only instances for which the five methods bring about different bundles were selected for further consideration. (In case of several tied winners, we have chosen a winner at random.) 

We used the following five aggregation methods: 

\begin{itemize}
  
  \item \textit{Maximizing utilitarian social welfare} (denoted \textsc{SUM}; see~\cite{abpb}). Namely, 

  \begin{equation}
  \begin{aligned}
    \textrm{argmax}_{B}\quad &\sum_{v_i\in V}\sum_{p\in B} v_i(p)\\
  \textrm{s.t.} \quad &\sum_{p\in B} C(p)\leq \ell\\
\end{aligned}
\end{equation}

  \item \textit{Maximizing Nash product} (denoted \textsc{NASH}; see~\cite{fairknapsack}). Namely,
  \begin{equation}
  \begin{aligned}
    \textrm{argmax}_{B}\quad &\prod_{v_i\in V}\sum_{p\in B} v_i(p)\\
  \textrm{s.t.} \quad &\sum_{p\in B} C(p)\leq \ell\\
\end{aligned}
\end{equation}
  
  \item \textit{Maximizing egalitarian social welfare} (denoted \textsc{EGAL}; see~\cite{aziz2018egalitarian}). Namely, 
    \begin{equation}
  \begin{aligned}
    \textrm{argmax}_{B}\quad &\min_{v_i\in V}\sum_{p\in B} v_i(p)\\
  \textrm{s.t.} \quad &\sum_{p\in B} C(p)\leq \ell\\
\end{aligned}
\end{equation}

  \item \textit{Minimal Transfers over
Costs} (denoted \textsc{MTC}; see~\cite{skowron2020participatory}). \\
  A recently proposed PB algorithm through which, conceptually, each voter is given one coin and specifies how this coin should be split among the projects. Then, iteratively, projects with sufficient funding are funded while others are eliminated from further consideration.
Indeed, in essence, MTC is an adaptation of STV to the setting of PB,\footnote{The ordinal, multiwinner version of STV basically works as follows: we start with an empty solution; in each iteration, if there is a candidate with at least (roughly) a $k$-fraction of the population ranking at the top, then we add it to the solution and remove these voters from further consideration; while if there is no such candidate, then we eliminate a candidate with the least number of voters ranking at the top, and reiterate.} so MTC operates in iterations, where in each iteration, if there is a project with sufficient voter support, then it is being funded, its support from the voters is being used, and the support's excess is being redistributed according to the supporting voters' ballots; otherwise, if there is no project with sufficient voter support, then a project that has the least chance of being supported, even after redistribution of voter support is being abandoned and its support being redistributed to our projects, again according to the supporting voters' ballots.
\new{Note that, while MTC can be used for cumulative ballots~\cite{skowron2020participatory}, it can be naturally used for approval ballots as well; in particular, in our experiments we have taken the $3$ projects with the highest cardinal utility as the approved projects. Since the participants were told that the utilities are additive and objective, we feel that this is a natural adaptation from cardinal ballots to approval ballots. Regarding the specific number of approvals, preliminary experiments show that there is not much of a difference between choosing any of $2$, $3$, or $4$-approval.}
  
\item \textit{Budgetary Proportional Justified Representation} (denoted \textsc{BPJR}; see~\cite{aziz2018proportionally}). \\
A recently proposed PB algorithm through which, each group of voters is \say{appropriately} represented in the solution bundle based on their cohesiveness (i.e., alignment with each other in terms of their preferences). In essence, this algorithm is designed so as to satisfy an adaptation of the multwinner axiom of Proportional Justified Representation~\cite{aziz2017justified} to the PB setting.
\new{In particular, we have adapted the algorithm presented in Proposition 3.7 by Aziz et al.~\cite{aussieone}; as we are dealing with cardinal utilities, while Aziz et al consider approval ballots, we consider the $3$ highest-utility projects as being approved by the voter (similarly to the way we adapt MTC to the approval case). Given such approval ballots, the algorithm we have implemented works as follows (the algorithm is of super-polynomial complexity, however it is feasible for our relatively small scale examples):
  we proceed in iterations, in which we start with an empty winning bundle that we eventually populate, and we consider a parameter $\ell$ that initially equals the total budget limit; initially we consider all bundles whose total cost equals $\ell$; among these, we choose the one for which the highest number of voters approve at least one of the projects in the bundle; then, we add all projects from this bundle to the winning bundle we populate and eliminate all voters who approve at least one of these projects from further consideration; after we are done with all bundles of total cost $\ell$, we decrease $\ell$ by one, and reiterate. As shown by Aziz et al, this algorithm indeed satisfied the proportionality axiom referred there as BPJR-L.}
  
  \end{itemize}

All aggregation methods discussed are proven to produce at least one bundle. When an aggregation method produces more than a single  bundle, one of them is chosen arbitrarily.

Prior to the presentation of the five instances (the instances appear in Appendix~\ref{app:exp1}), participants were asked to provide their age, gender, and economical status on a 5-point Likert scale, ranging from \textit{significantly above average} to \textit{significantly below average}. 
Then, the selected five instances were presented to the participants, in a random order, along with the motivational scenario discussed before. Participants were asked to assume the role of Alice (the manager of the shopping mall) and choose the most appropriate bundle of projects to fund. 
\new{Initially, we have asked participants to choose the most appropriate bundle to fund without any help from our end. Unfortunately, we quickly realized that the task was too time consuming and technical for most participants to properly complete.} 
To simplify the process, each instance was accompanied by the five output bundles calculated using the aggregation methods above, in a random order as well, from which each participant had to choose the most appropriate one to fund (as she sees fit). 

Furthermore, following the five instances, participants were asked to rank the following criteria from most to least important in their selections: (1) ``Maximize the expected revenue boost of the mall (summed over the stores)''; (2) ``Equalizing the revenue boost among the stores (roughly balancing the revenue boost across stores)''; and (3) ``Avoiding neglecting a store (maximize the revenue of the worst store)''. These three criteria were designed to roughly correspond with the first three aggregation methods discussed above (\textsc{SUM}, \textsc{NASH} and \textsc{EGAL}, respectively), allowing us to examine whether the provided rankings align with the participants' choices (we did not add verbal explanations corresponding to BPJR and MTC as these rules seem fairly hard to describe informally). Specifically, we are interested in examining  whether participants are individually-aligned in terms of the actual solution bundles they select and their ranking over the verbal descriptions of the corresponding aggregation rules. \new{The specific wording of the explanations were derived through a preliminary examination with our students of a few alternative options. Through this examination, we have made sure that the final wording, albeit imperfect as discussed later in this article, are correctly interpreted by potential participants.}

The survey, available in its English format in Appendix~\ref{app:exp1}, was administered during the months of July and August 2020 to two participant groups: 1) Israeli University students; and 2) Polish University students. Both groups were recruited by posting ads on computational courses' webpages in computer science and industrial engineering, offering them a chance of winning one of three gift-cards, each of 100NIS ($\sim$\$30), in a raffle. 
The Israeli group consists of $60$ participants, all of whom are students of Bar-Ilan University or Ben-Gurion University; $35$ male (and $25$ female); with an average age of $26.7$ (std = $7.03$). The Polish group consists of $40$ participants, all of whom are students of Jagelonnian University (Krakow); $29$ of which are male (and $11$ female); with an average age of $23.1$ (std = $5.2$).

\subsection{Experiment 1: Results}

\paragraph{Preprocessing.}
We start our analysis by omitting all answers of participants who completed the survey in an unreasonable time (in particular, in less than $5$ minutes) as to avoid possibly under-quality responses. Fortunately, only very few participants were omitted in this phase ($3$ from the Israeli group and $2$ from the Polish group). 


\paragraph{Chosen Bundles.}
Effectively, each participant has selected only one \say{most appropriate} bundle in each of the five presented instances. Through that selection, the participant has implicitly indicated which aggregation method had brought about the most appropriate bundle (according to her taste). We summed the number of times each aggregation method was chosen by the participants of each participant group. The results are presented in Table \ref{table:Exp1Selection}.

Starting with the Israeli group, using the Friedman test followed by posthoc Wilcoxon signed-rank test\footnote{The use of Friedman and Wilcoxon tests is due since normality cannot be adequately assumed for the collected data.} with Bonferroni correction,\footnote{Throughout this study, we correct for the specific multiple comparisons made in each analysis Bonferroni was used. We  have also tried using less conservative corrections such as Benjamini-Hochberg. Unfortunately, no significant changes was observed.} reveals that the \textsc{NASH} method was chosen significantly more often than any other aggregation method except for the \textsc{SUM} method, $p<0.05$. On the other hand, the \textsc{MTC} and \textsc{BPJR} methods were chosen significantly \textit{less} often than any other selection method, with $p<0.05$.
In other words, while we cannot conclude whether the popularity of the \textsc{SUM} method exceeds that of the \textsc{NASH} method, or vice versa, in a statistically significant manner, we can conclude that the \textsc{MTC} method and the \textsc{BPJR} method are significantly chosen less often than all the other three methods, including the \textsc{EGAL} method. The latter is statistically chosen less often than the \textsc{NASH} method.  
Similar results are obtained in the Polish group: the \textsc{SUM} and \textsc{NASH} methods are found to be chosen significantly more often than all other methods, $p<0.05$, yet \textsc{NASH} is chosen more often than \textsc{SUM} only at $p<0.1$. As with the Israeli group, the \textsc{CSTV} and \textsc{BPJR} methods are selected significantly less often than the other three methods.

	\begin{table} \centering
	\begin{tabular}{c | l |l }
		Method &  Israel & Poland\\ [0.5ex]
		\hline
		\hline
				\textsc{NASH}& \textbf{37.1\%} & \textbf{40.8\%}\\
		
	\textsc{SUM}& \textbf{30.7\%} & 29.2\%\\

		\textsc{EGAL}& 20.5\% & 17.4\%\\
		\textsc{MTC}& 7.9\% & 5.7\%\\
		\textsc{BPJR}& 3.8\% & 6.2\%\\
		\hline
	\end{tabular}
		\caption{Results of Experiment 1:
		The table shows the distribution regarding the aggregation method selected by participants (rows) divided by country (columns). Columns do not necessarily sum up to 1 due to rounding. Numbers in bold are discussed in the main text.}
	\label{table:Exp1Selection}
	\end{table}

\paragraph{Individual Consistency.}
We further examine the participants' \textit{individual consistency}: that is, we look at how consistent participants were in their five selections, in terms of the preferred aggregation methods.
%
In particular, we say that a participant is \textit{inconsistent} if no more than 2 of her selections correspond to the same aggregation method. On the other hand, we say that a participant is \textit{reasonably consistent} if 3 of her selections coincide on the same method. A participant is considered \textit{consistent} if at least~4 of her selections coincide on the same method. For both the Israeli and Polish groups, about 40\% of the participants were consistent, an additional 40\% were reasonably consistent, while only less than 20\% were inconsistent; see Table~\ref{table:Exp1Consis}.

	\begin{table} \centering
	\begin{tabular}{c | l |l }
		Consistency &  Israel & Poland\\ [0.5ex]
		\hline
		\hline
		\textsc{5}& 19.3\% & 21.6\%\\
		\textsc{4}& 19.3\% & 27.1\%\\
		\textsc{3}& 42.1\% & 37.8\%\\
		\textsc{2}& 19.3\% & 16.2\%\\
    	\textsc{1}& -\% & 2.7\%\\
		\hline
	\end{tabular}
		\caption{Results of experiment 1: Participants' individual consistency in their bundle selections. Rows denote the maximal number of  participant's selections, which coincide on the same aggregation method divided by country (columns). For example, in the Israeli participant group, $42.1\%$ of the participants chose exactly three bundles (out of 5) that coincide on the same aggregation method. Columns do not necessarily sum up to 1 due to rounding.}
	\label{table:Exp1Consis}
	\end{table}

For both groups, no significant differences were found between men and women nor between the participants' stated economical statuses. Also, similar results are observed when the Israeli and Polish participants are grouped together.

\paragraph{Ranked Explanations.}
%
%
We turn to analyze the participants' rankings over the three verbal descriptions, which correspond to the \textsc{SUM}, \textsc{NASH}, and \textsc{EGAL} aggregation methods. For simplicity, we overload the name of the aggregation method to denote its verbal explanation as well. Using Friedman's test, followed by posthoc Wilcoxon signed-rank test with Bonferroni correction, we find that, to the contrary to what we initially expected, \textsc{SUM} was ranked significantly higher than the other two criteria, $p<0.05$. In the Israeli group, \textsc{SUM} averaged a rank of 1.45 compared to average ranks of 2.2  and 2.4 achieved by \textsc{NASH} and \textsc{EGAL}, respectively. Similarly, in the Polish group, \textsc{SUM} averaged a rank of 1.5 compared to average ranks of 2.3  for both \textsc{NASH} and \textsc{EGAL}. Despite the significant popularity of the \textsc{NASH} method in the participants' bundle selections (as discussed above), no significant differences were found between \textsc{NASH} and \textsc{EGAL} in their rankings. The results are also presented in Table \ref{table:Exp1Expl}. 

Recall that no significant differences were found between men and women in terms of their aggregation method choices. However, we did find significant differences in terms of their provided rankings over the three criteria. When grouping the Israeli and Polish participants together by gender we find that men are significantly more likely to rank \textsc{SUM} higher than \textsc{NASH} and \textsc{EGAL} in their rankings while no significant differences are encountered in the female group, $p<0.05$. The difference is found to be significant only when combining the two groups. When comparing the rankings of men to that of women, using Mann-Whitney U-test we find that women are significantly more likely to rank \textsc{NASH} higher than men rank it, yet no other differences are found to be statistically significant, $p<0.05$. 
As was the case in the participants' bundle selections, no significant differences were encountered based on the participants' stated economical status.

Finally, we compare participants' aggregation method choices with their ranking over the three criteria. To that end, we compare the participant's provided ranking over the three criteria with the implicit ranking provided by her choices. Specifically, the \textsc{SUM}, \textsc{NASH} and \textsc{EGAL} aggregation methods were ranked by their recurrence in each participant's choices. Since two or more of the methods could be ranked the same (we are ignoring \textsc{BPJR} and \textsc{MTC} choices in this analysis), each method was assigned a possibly non-unique rank between 1 and 3 resulting in a 3-valued vector. For example, in the case where a participant chose \textsc{SUM} and \textsc{NASH} twice and \textsc{EGAL} once, her vector would read $[1,1,3]$, respectively. This vector was then compared with the actual ranking provided by that participant. As could be expected from the results described above, the two only poorly align. Specifically, considering an $L_1$ distance function, for only 14 Israeli participants (23\%) and 7  Polish participants (17.5\%) was the distances less or equal to one. For half of the Israeli participants and 45\% of the Polish participants the distance was greater than 2. Since $L_1$ may introduce some shortcomings, we have also examined other similarity measures such as the $L_2$ distance function or cosine similarity which brought about similar results.   

	\begin{table} \centering
	\begin{tabular}{c | l |l }
		Explanation &  Israel & Poland\\ [0.5ex]
		\hline
		\hline
				\textsc{NASH}& 2.2 & 2.3\\
		
	\textsc{SUM}& \textbf{1.45} & \textbf{1.5}\%\\

		\textsc{EGAL}& 2.5\% & 2.3\%\\
		\hline
	\end{tabular}
		\caption{Results of Experiment 1:
		The table shows the  participants' average rankings over the three verbal descriptions, which correspond to the \textsc{SUM}, \textsc{NASH}, and \textsc{EGAL} aggregation methods. The verbal descriptions ranked by participants (rows) are divided by country (columns). Numbers in bold are discussed in the main text.}
	\label{table:Exp1Expl}
	\end{table}

\subsection{Experiment 1: Discussion}

There are two central results in Experiment 1: first, the \textsc{NASH} aggregation method is the one deemed most appropriate \textit{by observing the bundles chosen by the participants}; Second, the \textsc{SUM} aggregation method is the one deemed most appropriate \textit{by observing the verbal explanations ranked by the participants}.
These results are surprising as the \textsc{NASH} and \textsc{SUM} aggregation methods are not usually adopted by researchers and practitioners in real-world PB elections (although a greedy approximation to \textsc{SUM} is the most popular real-world aggregation method in use). 

The possible discrepancy between the two results may be due to the specific verbal explanation chosen. Indeed, while we tried to be objective and clear in our verbal descriptions, we acknowledge that perhaps we were not successful in this task.  
In this context, it is important to mention the recent work on \emph{explainable social choice} (see, e.g.,~\cite{peters2020explainable}), as it is indeed of great value to identify useful ways to explain aggregation methods to common people (as well as to examine how different explanations may influence perceptions and decisions).
This fact, combined with the fact that participants were rather consistent in their aggregation method choices suggest that \textsc{NASH} is probably the most preferred method in our participant groups. 

Generally, demographic factors were not found to bear significant importance in our experiment. One exception is the fact that men tend to rate the verbal explanation of \textsc{SUM} higher than any other aggregation method and higher than how female participants rate it.

Overall, we find strong evidence to support the appropriateness of the \textsc{NASH} aggregation method across our participants and scenarios as compared to the other examined methods. In addition, we identify some support for the \textsc{SUM} method as well, largely due to the participants' rankings over the verbal explanations.

\section{Experiment 2: Approval Ballots}\label{section:experiment2}

Experiment 1 (Section \ref{section:experiment1}) has focused on identifying which aggregation method is preferred by common people, given voters' \textit{cardinal utilities}. As discussed before, in many real-world settings, obtaining cardinal utilities is impractical or infeasible (also, currently most real-world PB processes use approval ballots and not cardinal utilities; note that other ballot type exist, e.g., Knapsack ballots~\cite{goel2019knapsack}, however we have chosen approval ballots for their simplicity and popularity). 
Thus, in our second experiment we turn to consider the case where explicit, cardinal utilities are unavailable. Instead, we assume that each voter provides us with a subset of \say{approved} projects $v_i$ such that $p\in v_i$ if voter $i$ approves project $p$. \new{Indeed, this is the setting of approval-based PB~\cite{abpb}.} 
We focus on the \textsc{SUM} and \textsc{NASH} aggregation methods that, according the results of Experiment~1, favorably compare to the other examined competing methods. Specifically, we seek to investigate how non-experts translate voter preferences, as expressed by approval ballots, into cardinal utilities that can, in turn, be aggregated using either \textsc{SUM} or \textsc{NASH}.

To this end, as was the case in Experiment 1, we devise a simple yet natural PB scenario in which participants are tasked with choosing the most appropriate bundle to fund. We use the following motivational scenario in our experiment\footnote{Note that the participants are not aware whether we use SUM or NASH.}:

\smallskip\noindent\textit{\textbf{Motivating Scenario (Approval ballots):}
Alice is the owner of a company that manages a small residential building consisting of $N$ apartments.
Near the end of the fiscal year, NIS 50K were left in the building's account. She wishes to use the money to improve the quality of life for the residents.
For simplicity, assume that Alice cannot invest more than NIS 50K and assume further that unused money can not be used in the future.
Alice was offered $M$ possible projects (such as, enlarging the parking space, adding an elevator, etc.), where each project has a different cost.
Alice's job is to collect the residents' votes, where each vote corresponds to a set of approved projects, and choose which projects to fund.
\new{Indeed, the participants were presented with concrete values of $N$ and $M$.}
}
\smallskip

The above scenario was devised following this rationale: by using generic projects  we seek to avoid participants casting their own preferences into the decision setting. In addition, since Alice's utility is not defined at all (she is only introduced as the owner of the management company), we leave it up to the participants to decide what she constitutes as appropriate in this generic setting without introducing potentially biasing terms such as \say{fairness}. 

\subsection{Experiment 2: Design}

Following the motivational scenario outlined above, we first randomly generated a set of instances with $N$ voters (randomly chosen between 3 and 6), $M$ projects (randomly chosen between 5 and 7), voter approval ballots that were randomly generated such that each voter approved either $2$ or $3$ projects, and projects' costs, which were randomly drawn from a uniform distribution over \{1,2,\ldots,50\}. 
In order to examine different possible utility functions, each generated instance was first solved assuming five different utility functions using either \textsc{SUM} or \textsc{NASH} aggregation methods, as detailed next. Then, five instances for which none of the five output bundles coincide were selected for each aggregation method. Namely, 5 instances were selected for the \textsc{SUM} aggregation method and 5 instances were selected for the \textsc{NASH} aggregation method.

We used the following five utility functions to translate an approval ballot and a funded bundle into cardinal utilities (note that $v_i$ is used to denote an approval ballot on the right hand side and cardinal utilities on the left hand side): 

  \begin{itemize}
  \item \textit{Dichotomous utility} (denoted \textsc{0/1 utility}). Namely, 
  \[
  v_i(B) =  \mathbbm{1}_{v_i\cap B\neq\emptyset}
  \]
    
  \item \textit{Number of projects approved and funded} (denoted \textsc{\#Projects} utility). Namely,
  \[
  v_i(B) = \sum_{p\in B} \mathbbm{1}_{p \in v_i}
  \]
  
  \item \textit{Sum of costs of project approved and funded} (denoted \textsc{TotalCost} utility). Namely, 
  \[
  v_i(B) = \sum_{p\in v_i\cap B} C(p)
  \]

  \item \textit{Square root of the sum of costs of project approved and funded} (denoted \textsc{SqrtCost} utility). Namely, 
    \[
  v_i(B) = \sqrt{\sum_{p\in v_i\cap B} {C(p)}}
  \]

  \item \textit{Cost of most expensive project approved and funded} (denoted \textsc{MaxSet} utility). Namely,
    \[
  v_i(B) = \text{max}_{p\in v_i\cap B} {C(p)}
  \]
  
  \end{itemize}

When one of the above functions produces more than a single bundle, one of which is chosen arbitrarily. 

The five utility functions described above were devised based, in part, on the assumption that the cost of each funded project may play an important part in the way people estimate the associated utility derived from it. 

Since we assume 2 possible aggregation methods, the experiment is conducted in two separate phases: \textit{Experiment 2-\textsc{SUM}}, which considers the instances under the \textsc{SUM} aggregation method; and \textit{Experiment 2-\textsc{NASH}}, which considers the instances under the \textsc{NASH} aggregation method. \new{In the latter case, in order to avoid a Nash product of zero, all utilities were smoothed by adjusting the initial utility each voter receives to be $0.01$. }

In both phases, prior to presentation of the five instances (the instances appear in Appendix~\ref{app:exp2} in Hebrew but they are easily understandable in any language), participants were asked to provide their age, gender, and economical status on a 5-point Likert scale ranging from \textit{significantly above average} to \textit{significantly below average}. 
Then, the appropriate five instances were presented to participants in a random order, along with the motivational scenario discussed above. Participants were asked to assume the role of Alice (the owner of the management company) and choose the most appropriate bundle to fund. \new{Similarly to Experiment 1, here too, we have initially asked participants to choose the most appropriate bundle to fund without any help from our end. Unfortunately, once again, participants found the task too difficult, resulting in many drop-outs, poor quality answers and negative informal feedback. As such, we decided } to simplify the process such that each instance was accompanied by the five output bundles calculated using the appropriate utility function discussed above, in a random order as well, from which each participant had to choose the most appropriate one to fund (as she sees fit). 

Following the five instances, subjects were asked to rank the following criteria from most to least important in their selections: 
(1) Every voter should get at least one of her approved projects funded (if possible);
(2) The most approved projects should be funded;
(3) Approved expensive projects should be preferred over approved cheap projects;
(4) Projects funded and approved by every voter should cost roughly the same; and
(5) Every voter should get at least one \say{expensive} project funded from her approval ballot.  
These five criteria were designed to roughly correspond with the five methods discussed above (\textsc{0/1}, \textsc{\#Projects}, \textsc{TotalCost}, \textsc{SqrtCost} and \textsc{MaxSet}, respectively), allowing us to examine whether the provided ranking aligns with the participants' bundle choices; that is, whether participants are individually-aligned with the actual bundles they select and how they rank the verbal descriptions of the corresponding utility functions.

The two surveys, one assuming the \textsc{SUM} aggregation method and the other assuming the \textsc{NASH} aggregation method (both available in their Hebrew format in Appendix~\ref{app:exp2}), were administered during the months of September and October 2020 to two separate groups of 40 Israeli University students each. Both groups were recruited by posting ads on computational courses' webpages in computer science and information science, offering them a chance of winning one of three gift-cards, each of 100NIS, in a raffle ($\sim$\$30). 
All students are of Bar-Ilan University; 37 male; with an average age of 30.7 (std=9.76). The students were pseudo-randomly assigned to the two groups with no significant differences in age between the groups.

\subsection{Experiment 2: Results}



\paragraph{Preprocessing}
We start our analysis by omitting all answers of participants who completed the survey in an unreasonable time (in particular, less than 5 minutes) as to avoid possibly under-quality responses. Fortunately, only 5 participants were omitted in this phase from both groups combined.


\paragraph{Chosen bundles}
Effectively, each participant has selected only one \say{most appropriate} bundle in each of the five instances. Through that selection, the participant has implicitly indicated which utility method had brought about the most appropriate bundle in her opinion (under the examined aggregation method). We summed the number of times each aggregation method was chosen by the participants in each experiment (\textsc{SUM} and \textsc{NASH}). The results are presented in Table~\ref{table:Exp2Selection}.

Starting with the \textsc{SUM} experiment, using the Friedman test followed by posthoc Wilcoxon signed-rank test with Bonferroni correction, reveals that the \textsc{0/1} and \textsc{\#Projects} utility functions were chosen significantly more often than any other examined function, $p<0.05$. On the other hand, no statistically significant difference is found between them. The three remaining functions (\textsc{TotalCost}, \textsc{SqrtCost}, and \textsc{MaxSet}) display very low popularity with no significant differences between them either. In other words, while we cannot conclude whether the popularity of the \textsc{0/1} or \textsc{\#Projects} method is superior to the other in a statistically significant manner, we can conclude that both are significantly chosen more often than any of the other three functions. 

Interestingly, slightly different results are encountered for the \textsc{NASH} experiment. As was the case under the \textsc{SUM} experiment, using the Friedman test followed by posthoc Wilcoxon signed-rank test with Bonferroni correction, reveals that \textsc{\#Projects} is significantly chosen more often than any other function examined, $p<0.05$. Surprisingly,  to the contrary of its popularity in the \textsc{SUM} experiment, the \textsc{0/1} method is significantly outchosen by all other examined functions, $p<0.05$. In addition, \textsc{TotalCost} is found to be chosen significantly more often than \textsc{SetMax}, $p<0.05$. All other differences were not found to be statistically significant. In other words, the \textsc{\#Projects} function is significantly more popular in comparison to the other examined functions, while the \textsc{0/1} function is found to be the least popular in a statistically significant manner.

	\begin{table} \centering
	\begin{tabular}{c | l |l }
		Utility &  \textsc{SUM} & \textsc{NASH}\\ [0.5ex]
		\hline
		\hline
		\textsc{0/1}& \textbf{40.5\%} & 5.8\% \\
		\textsc{\#Projects}& \textbf{49\%}& \textbf{38.3\%} \\
		\textsc{TotalCost}& 3.3\% & 21.6\% \\
		\textsc{SqrtCost}& 5.6\% & 19.2\%\\
		\textsc{MaxSet}& 2\% & 15\% \\
		\hline
	\end{tabular}
		\caption{Results of Experiment 2:
		The table shows the distribution regarding the utility function selected by participants (rows) divided by the aggregation method used in the experimental setting (rows). Columns do not necessarily sum up to 1 due to rounding. Results in bold are discussed in the text.}
	\label{table:Exp2Selection}
	\end{table}

\paragraph{Individual Consistency.}
We further examine the participants' \textit{individual consistency}: specifically, we look at how consistent participants were in their five selections in terms of the selected utility functions. 
As in Experiment 1, we say that a participant is \textit{inconsistent} if no more than 2 of her selections correspond to the same utility function. On the other hand, we say that a participant is \textit{reasonably consistent} if 3 of her selections coincide on the same utility function. A participant is considered \textit{consistent} if at least~4 of her selections coincide on the same function. Similar to the results presented in Table \ref{table:Exp1Selection} we see that under the \textsc{SUM} experiment, 41.8\% of the participants were consistent, an additional 34.2\% were reasonably consistent and only 24\% were inconsistent. On the other hand, under the \textsc{NASH} experiment, only 10.9\% of the participants were consistent, an additional 29.8\% were reasonably consistent and 59.6\% were inconsistent; see Table~\ref{table:Exp2Consis}.

	\begin{table} \centering
	\begin{tabular}{c | l |l }
		Consistency &  SUM & NASH\\ [0.5ex]
		\hline
		\hline
		\textsc{5}& 18.2\% & 2.7\%\\
		\textsc{4}& 23.6\% & 8.2\%\\
		\textsc{3}& 34.2\% & 29.8\%\\
		\textsc{2}& 24\% & 54.1\%\\
    	\textsc{1}& -\% & 5.5\%\\
		\hline
	\end{tabular}
		\caption{Results of Experiment 2: Participants' individual consistency in their selected utility functions. Rows denote the maximal number of  participant's selections, which coincide on the same utility function divided by aggregation method used (columns). Columns do not necessarily sum up to 1 due to rounding.}
	\label{table:Exp2Consis}
	\end{table}

We now turn to examine potential differences between men and women. Starting with the \textsc{SUM} experiment, using Fisher's exact text we observe that men and women choose differently, $p<0.05$. Indeed, while roughly 10\% of men and women's selections were either \textsc{TotalCost}, \textsc{SqrtCost}, or \textsc{MaxSet}, a notable difference was found in their selection of \textsc{0/1} and \textsc{\#Projects}. Specifically, about 70\% of men choices were \textsc{0/1}, while about 70\% of women choices were \textsc{\#Projects}. In other words, there is a significant difference in the way men and women select their preferred utility method -- men seem to prefer \textsc{0/1} whereas women seem to prefer \textsc{\#Projects}. Interestingly, no significant differences were found between men and women under the \textsc{NASH} experiment. This may be partially attributed to the general inconsistency in the participants' selection under the \textsc{NASH} experiment, as shown in Table \ref{table:Exp2Consis}.

As is the case for gender-based differences, we find significant differences between self-reported economical statuses under the \textsc{SUM} experiment. Using a Chi-square test followed by post-hoc pairwise Chi-square tests with Bonforroni corrections we find that participants who consider their economical status below average (or very below average) choose the \textsc{\#Projects} significantly more often than the others. About 66\% of the former's choices were \textsc{\#Projects} while 62\% of the latter's choices were \textsc{0/1}. As before, no significant differences were found under the \textsc{NASH} experiment.

\paragraph{Ranked explanations}
We turn to analyze the participants' rankings over the five utility criteria, which correspond to the \textsc{0/1}, \textsc{\#Projects}, \textsc{TotalCost}, \textsc{SqrtCost}, and \textsc{MaxSet} utility functions. For simplicity, we overload the name of the utility function to denote the criteria as well. Surprisingly, we find that, under both the \textsc{SUM} and \textsc{NASH} experiments, participants ranked \textsc{0/1} and \textsc{\#Projects} significantly higher than the other three examined criteria, $p<0.05$. 

Specifically, under the \textsc{SUM} experiment,  \textsc{0/1} and \textsc{\#Projects} averaged ranks of 1.5 and 1.65, respectively, while the remaining three averaged between 3.6 and 4.2, with the \textsc{MaxSet} criteria ranking  significantly lower than all other criteria (averaging 4.2). Interestingly, slightly more than half of the participants ranked \textsc{0/1} at number one while the remaining (slightly less than) half ranked \textsc{\#Projects} at the first position. None ranked any of the remaining three criteria at the first position. 

Similarly, under the \textsc{NASH} experiment, \textsc{0/1} and \textsc{\#Projects} averaged ranks of 1.8 and 1.75, respectively, while the remaining three averaged between 3.3 and 4.6, with the \textsc{MaxSet} criteria ranking significantly lower than all other criteria (averaging 4.6). Slightly less than half of the participants ranked \textsc{0/1} at the first position while the remaining (slightly more than) half ranked \textsc{\#Projects} at the first position. Only $2$ participants ranked \textsc{TotalCost} at the first position under this experiment.  

The results are summarized in Table \ref{table:Exp2Expl}.

When grouping \textsc{SUM} and \textsc{NASH} participants together by gender, we find that men are significantly more likely to rank \textsc{0/1} higher than any other criteria in their rankings while female participants are significantly more likely to rank \textsc{\#Projects} higher than men. No significant differences were encountered based on the participants' stated economical status.

Finally, we compare the participants' utility function choices with their ranking over the five verbal explanations. To this end, we use the same vector-based approach used in Experiment 1 (Section \ref{section:experiment1}), where participants' \say{implicit} ranking over the utility methods is compared to their explicit ranking over the corresponding verbal explanations. Considering an $L_1$ distance function, under the \textsc{SUM} and \textsc{NASH} experiments, for most participants ($66\%$ and $54\%$, respectively) the distances were greater than $4$, and for the larger part of the remaining participants ($28\%$ and $29\%$, respectively), the distances were $3$ or $4$. Other similarity measures such as the $L_2$ distance function or cosine similarity brought about similar results.
While the similarity measurements described above display rather poor alignment between the participants' bundle choices and rankings over verbal descriptions, when examining only the top ranked verbal descriptions provided by each participant we find an interesting phenomena: under the \textsc{SUM} experiment: for $83\%$ of the participants the top ranked criteria was ranked among the top \textit{two} in their utility function choices. This was the case for only $54\%$ of the participants under the \textsc{NASH} experiment. Using this similarity perspective, participants' top ranked verbal explanation was generally compatible with their top utility choices, especially under the \textsc{SUM} experiment.

	\begin{table} \centering
	\begin{tabular}{c | l |l }
		Explanation &  SUM & NASH\\ [0.5ex]
		\hline
		\hline
		\textsc{0/1}& \textbf{1.5} & \textbf{1.8} \\
		\textsc{\#Projects}& \textbf{1.65}& \textbf{1.75} \\
		\textsc{TotalCost}& 3.6\% & 3.3\% \\
		\textsc{SqrtCost}& 3.8\% & 3.65\%\\
		\textsc{MaxSet}& 4.2\% & 4.6\% \\
		\hline
	\end{tabular}
		\caption{Results of Experiment 2:
		The table shows the  participants' average rankings over the five utility criteria, which correspond to the \textsc{0/1}, \textsc{\#Projects}, \textsc{TotalCost}, \textsc{SqrtCost}, and \textsc{MaxSet} utility functions. The verbal descriptions ranked by participants (rows) are divided by country (columns). Numbers in bold are discussed in the main text.}
	\label{table:Exp2Expl}
	\end{table}

\subsection{Experiment 2: Discussion}

The central result of Experiment 2 is that the \textsc{\#Projects} utility function stands out under both the \textsc{SUM} and \textsc{NASH} aggregation methods \textit{by observing both the bundles chosen by the participants as well as the participants' provided rankings over the verbal descriptions}. This suggests that \textsc{\#Projects} is the utility function deemed most appropriate by our participants. The \textsc{0/1} utility function is also popular, mainly under the \textsc{SUM} aggregation method experiment and with respect to the participants' rankings over the verbal explanations. 
To us, these two results are very surprising as we have initially speculated that any reasonable utility function should account for the projects' costs. 
The popularity of the \textsc{\#Projects} utility function in the bundle selections of the participants as well as its high position in their ranking over the verbal explanations \textit{under both aggregation methods} provide very strong evidence to support its appropriateness. Some support for the \textsc{0/1} utility function can also be identified as discussed before. 


It is important to note that participants were generally \textit{consistent} in their bundle selections under the  \textsc{SUM} aggregation method but \textit{inconsistent} under the \textsc{NASH} aggregation method.
We believe that this result should be attributed, in part, to the somewhat more complex mechanism of the \textsc{NASH} aggregation method compared to the simplistic \textsc{SUM} method. Specifically, while it is very easy to calculate the utility for each voter \textit{separately} given a potential utility function, it is much easier to calculate the summation over these utilities rather than their multiplication. This additional level of \say{complexity} may have introduced some noise into the results. 

Interestingly, unlike Experiment 1, demographic-based differences between participants were found under the \textsc{SUM} aggregation method but not under the \textsc{NASH} aggregation method. Since the results were not consistent across the two phases, we plan to continue this line of work in the future.   

\section{Experiment 3: Unconstrained Bundle Selection}

Note that in Experiments 1 and 2, we have explicitly encoded several aggregation methods in order to simplify the participants' task. However, this design may also introduce some bias to the results, specifically, participants could have chosen differently if they were to make their own budget-feasible bundles. 
In order to examine this issue we replicate Experiments 1 and 2, yet this time, we do not provide the participants' with pre-computed bundles which correspond to the examined aggregation methods. To that end, we recruited 35 new participants who have not taken part in Experiment 1 or 2. All participants are Israeli Master students who attend the authors' courses. 

In a preliminary investigation with volunteers from our labs we have found that replicating the two experiments without providing the pre-computed bundles was significantly more complex than we thought. Specifically, our volunteers indicated that solving more than very few instances at a time was simply too hard for them to do properly. As such, we have decided to randomly allocate 15 participants to the replication of Experiment 1, which we will refer to as Experiment 3.1, and 20 to the replication of Experiment 2, which we will refer to as Experiment 3.2. In both, we have asked participants to solve \say{at least three settings} out of the randomly ordered set. 

\subsection{Results}

Starting with Experiment 3.1, the 15 participants allocated to this experiment have provided 74 solutions in total (average of 5 solution per participant and 15 per setting). First, only 5\% of the provided solutions were infeasible, suggesting that the participants have understood their task. In addition, only 5\% have provided a bundle different from those calculated as part of Experiment 1. As can be observed from Table \ref{table:Exp31}, the results are very akin to those of Experiment 1 (Table \ref{table:Exp1Selection}). 

	\begin{table} \centering
	\begin{tabular}{c | l  }
		Method &  Israel \\ [0.5ex]
		\hline
		\hline
				\textsc{NASH}& 30\% \\
	\textsc{SUM}& 30\%\\
		\textsc{EGAL}& 15\%  \\
		\textsc{MTC}& 10\%  \\
		\textsc{BPJR}& 5\% \\
        \textbf{Infeasible}& \textbf{5\%}  \\
		\textbf{Unaccounted}& \textbf{5\%}  \\
		\hline
	\end{tabular}
		\caption{Results of Experiment 3.1:
		The table shows the distribution regarding the aggregation method selected by participants. Columns do not necessarily sum up to 1 due to rounding. Numbers in bold are discussed in the main text.}
	\label{table:Exp31}
	\end{table}

Turning to Experiment 3.2, the 20 participants allocated to this experiment have provided 162 solutions in total (average of 8 solution per participant and 16 per setting). First, only 5.7\% of the provided solutions were infeasible, suggesting that the participants have understood their task. In addition, only 2.5\% have provided a bundle different from those calculated as part of Experiment 2. As can be observed from Table \ref{table:Exp2Selection}, the results are very akin to those of Experiment 2 (Table \ref{table:Exp2Selectionb}). 

	\begin{table} \centering
	\begin{tabular}{c | l }
		Utility &  \textsc{Israel}\\ [0.5ex]
		\hline
		\hline
		\textsc{0/1}& 22.8\% \\
		\textsc{\#Projects}& 46.2\%\\
		\textsc{TotalCost}& 8.2\%  \\
		\textsc{SqrtCost}& 9.5\% \\
		\textsc{MaxSet}& 5\% \\
		\textbf{Infeasible}& \textbf{5.7\%}  \\
		\textbf{Unaccounted}& \textbf{2.5\%}  \\
		\hline
	\end{tabular}
		\caption{Results of Experiment 2:
		The table shows the distribution regarding the utility function selected by participants. Columns do not necessarily sum up to 1 due to rounding. Results in bold are discussed in the text.}
	\label{table:Exp2Selectionb}
	\end{table}

These results combine to suggest that our selected aggregation methods did not introduce a significant bias in Experiments 1 and 2.

\section{Conclusions and Outlook}

Through the above human study, we investigated what ordinary people (i.e., non-experts) deem as appropriate solutions to instances of participatory budgeting. Our study comprised of several artificially-generated instances of participatory budgeting for which non-experts were asked to select the bundles they consider to be the most appropriate. In addition, we have asked our participants to rank verbal explanations that correspond to different aggregation methods for participatory budgeting.
We focus on both the case of real-valued utilities (Experiment~1) as well as on the case of approval ballots (Experiment~2). 

The main conclusion from Experiment~1 is that most people select the \textsc{NASH} method (i.e., maximizing the Nash product of voter utilities) or the SUM method (i.e., maximizing the sum of voter utilities), with NASH being more popular than \textsc{SUM}. Our confidence in this conclusion is rather high, as participants were generally consistent in their choices. 

As the first experiment consisted of instances with real-valued utilities, the two most popular aggregation methods were selected for further investigation under the setting of approval ballots in Experiment~2. 
%
The main conclusions from Experiment~2 is that most people select bundles that correspond to the \textsc{\#Projects} utility function (i.e., the utility of a voter equals the number of projects approved by the voter that are being funded in the winning bundle). The second most popular utility function is the \textsc{0/1} function, that assumes a unit utility when at least one project approved by the voter is funded.
Here, as well, our confidence in the conclusions is rather high, as our participants were generally consistent in their choices.

Despite the seemingly coherent results discussed above with respect to the bundles selected by our participants, we have observed a discrepancy between the bundles participants select and their ranking of the corresponding verbal explanations. This discrepancy is more visible in Experiment~1 and when considering the \textsc{NASH} aggregation method in Experiment~2, but less visible when considering the \textsc{SUM} aggregation method in Experiment~2. We believe that this is at least partially due to the \textsc{NASH} method being a more complicated aggregation method. We view this issue as a good motivating example for the necessity of \emph{explainable social choice}~\cite{peters2020explainable}, i.e., the need to find good ways to communicate aggregation methods to common people. It is, indeed, important to note that these results might be also influenced by the specific verbal explanations we have chosen.

We recognize that the current study is limited by the amount, quality, and diversity of the data used. In the context of this work, our participant pool was neither very large nor very heterogeneous and consisted of 215 Israeli and Polish university students. This may hinder the generalization of our findings in the general population. Future replication of this study in the general population could address this concern. In addition, our PB settings were relatively small in terms of number of projects and voters compared to how PB is commonly practiced by municipalities. \new{It is, however, important to note that in a preliminary informal examination of the matter we have found that testing larger instances (e.g., $>10$ voters \new{or} projects) may be too complex for non-experts and is likely to bring about poor-quality answers. This result is also supported by a recent study \cite{laruelle2021voting}. As such, we plan to examine the issue of scalability in the future through other means such as in-depth interviews with potential decision makers.} 
It is also important to note that some technical study design decisions have also had an effect the results. For example, in Experiment 2, we assumed that voters approved their top 2 or top 3 project while, in some settings, voters could have approved a different number of projects. 
Furthermore, our random tie-breaking may indeed have an effect on the results, in particular wrt.\ the Nash product.
\new{Last, it is important to remember that different aggregation methods may provide the same solution for a given PB setting. In our empirical evaluation, we focus on settings for which the examined aggregation methods disagree. We found that any two methods we examined  provide different solutions for 19\%-42\% of the randomly generated PB settings we examined. }

We plan to extend this work in several directions:
  first, we seek to establish a collaboration with an Israeli municipality in order to examine our findings in a real-world large scale PB setting. Since the standard aggregation method used today is a greedy approximation algorithm to \textsc{SUM} over \textsc{\#Projects}, real-world results of additional methods could possibly shape the way PB is practiced. 
  Second, we plan to consider various ballot types, such as Knapsack votes.
  Third, we plan to investigate what non-experts deem desirable in additional non-trivial social choice settings such as mutliwinner elections~\cite{mwchapter}.
  \new{Last, we plan to further examine additional human-centered aspects of PB decision making such as the presumed need for transparency or explainability from an aggregation method to be useful in practice. Specifically, we wish to examine how PB outcomes should be best mitigated to human decision makers. }

\section*{Acknowledgements}

Nimrod Talmon was supported in part by the Israel Science Foundation (ISF; Grant No. 630/19).

\bibliographystyle{spmpsci}
\bibliography{bib}

\appendix

\section{Experiment 1: Instances}\label{app:exp1}

\begin{figure}[t]
\centering
\includegraphics[width=0.85\textwidth]{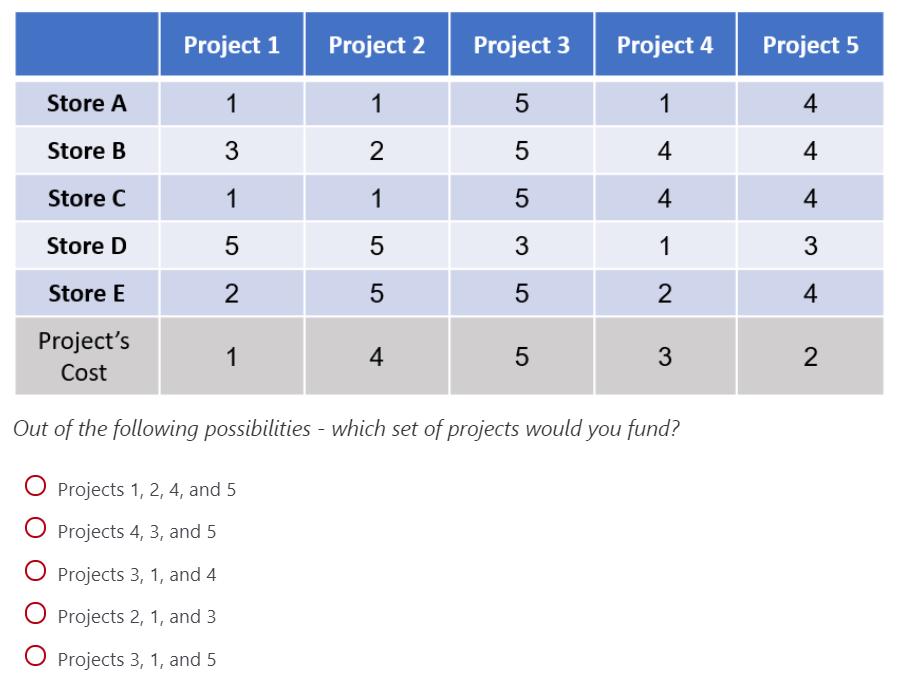}
\caption{Experiment 1: Scenario 1 out of 5}
\label{figure:oneone}
\end{figure}

\begin{figure}[t]
\centering
\includegraphics[width=0.85\textwidth]{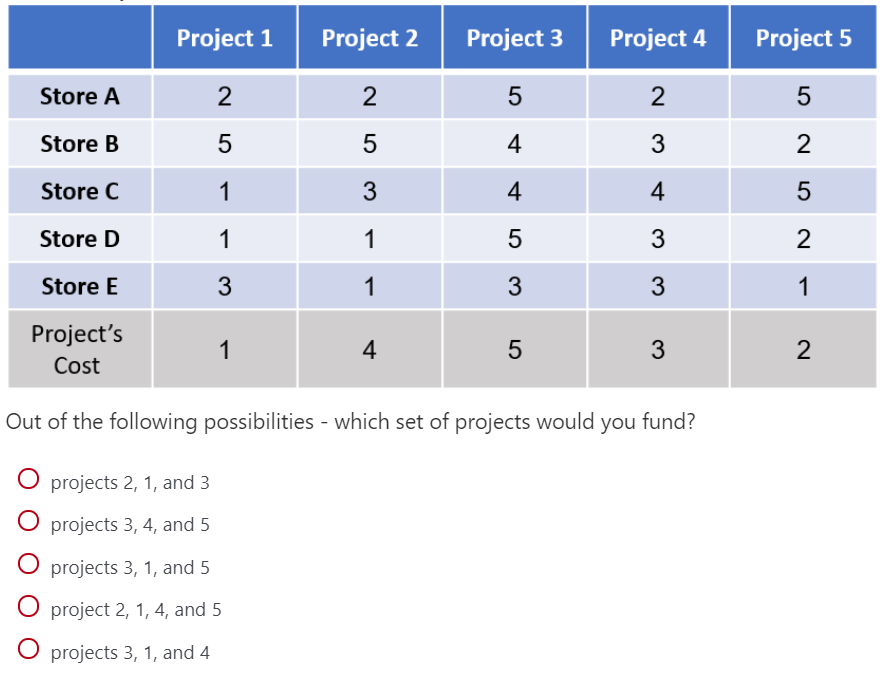}
\caption{Experiment 1: Scenario 2 out of 5}
\label{figure:onetwo}
\end{figure}

\begin{figure}[t]
\centering
\includegraphics[width=0.85\textwidth]{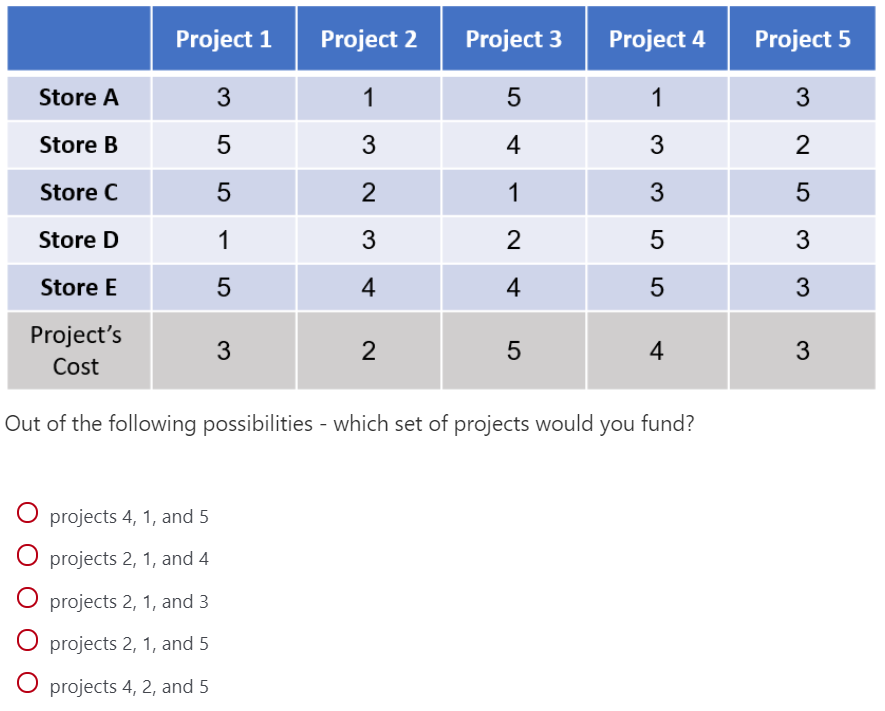}
\caption{Experiment 1: Scenario 3 out of 5}
\label{figure:onethree}
\end{figure}

\begin{figure}[t]
\centering
\includegraphics[width=0.85\textwidth]{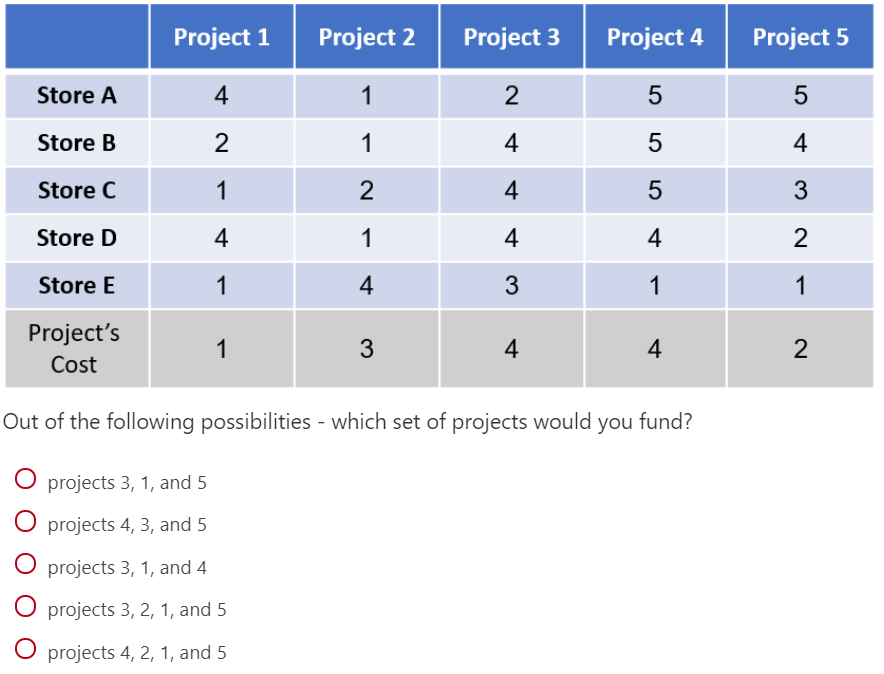}
\caption{Experiment 1: Scenario 4 out of 5}
\label{figure:onefour}
\end{figure}

\begin{figure}[t]
\centering
\includegraphics[width=0.85\textwidth]{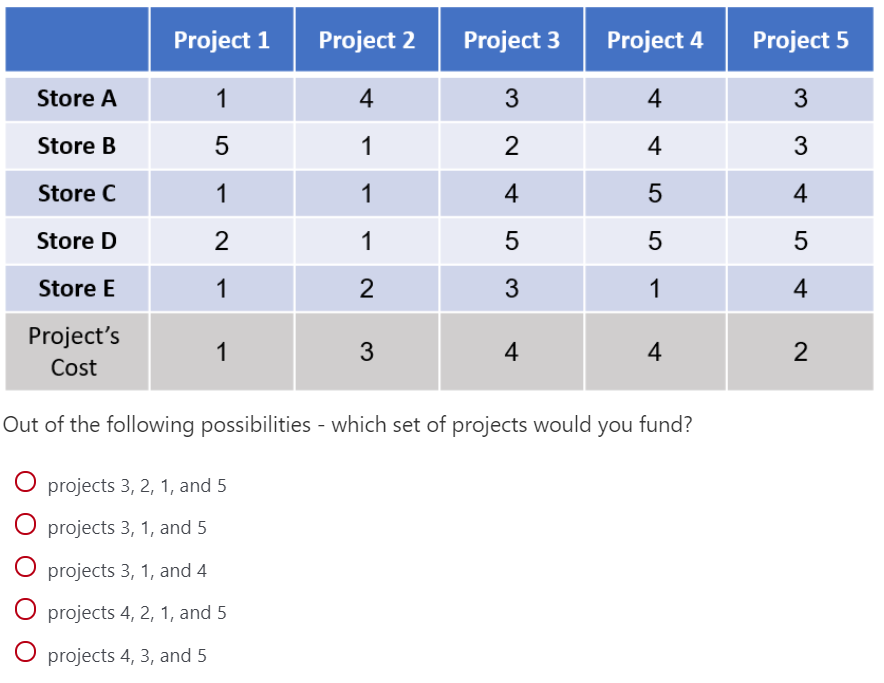}
\caption{Experiment 1: Scenario 5 out of 5}
\label{figure:onefive}
\end{figure}

The five instances used in Experiment 1 (see Section~\ref{section:experiment1}) appear in Figures~\ref{figure:oneone},~\ref{figure:onetwo},~\ref{figure:onethree},~\ref{figure:onefour}, and~\ref{figure:onefive}.

\section{Experiment 2: Instances}\label{app:exp2}

\begin{figure}[t]
\centering
\includegraphics[width=0.85\textwidth]{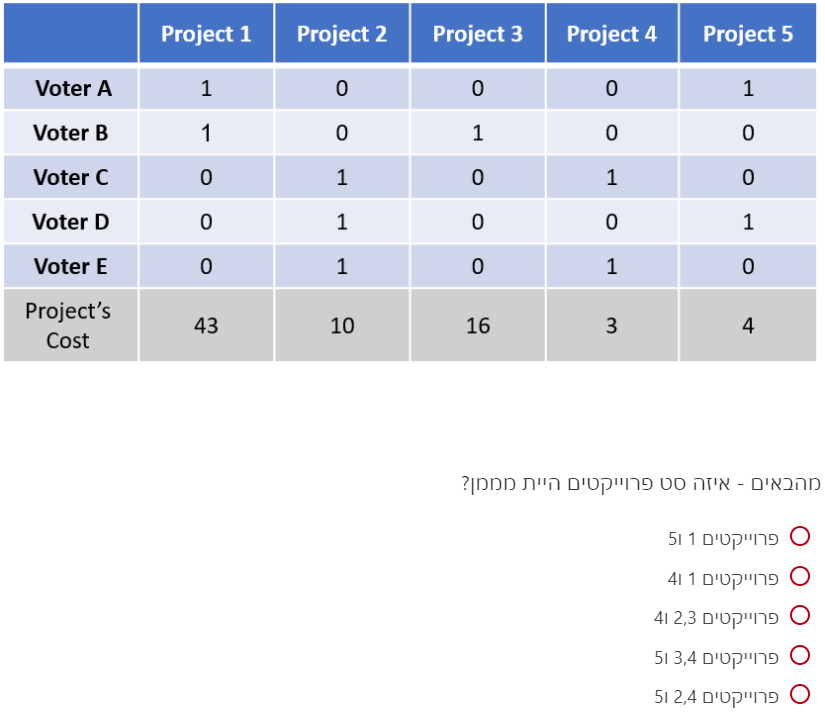}
\caption{Experiment 2-\textsc{SUM}: Scenario 1 out of 5}
\label{figure:twoone}
\end{figure}

\begin{figure}[t]
\centering
\includegraphics[width=0.85\textwidth]{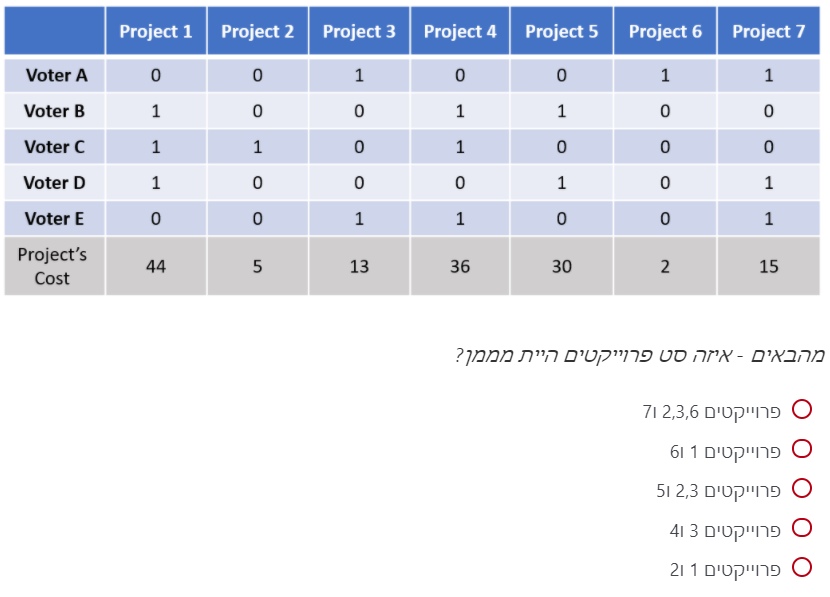}
\caption{Experiment 2-\textsc{SUM}: Scenario 2 out of 5}
\label{figure:twotwo}
\end{figure}

\begin{figure}[t]
\centering
\includegraphics[width=0.85\textwidth]{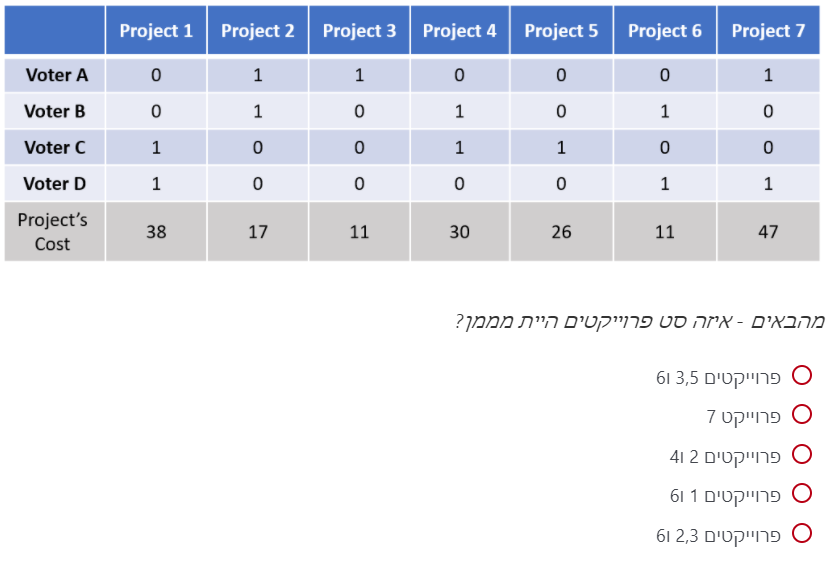}
\caption{Experiment 2-\textsc{SUM}: Scenario 3 out of 5}
\label{figure:twothree}
\end{figure}

\begin{figure}[t]
\centering
\includegraphics[width=0.85\textwidth]{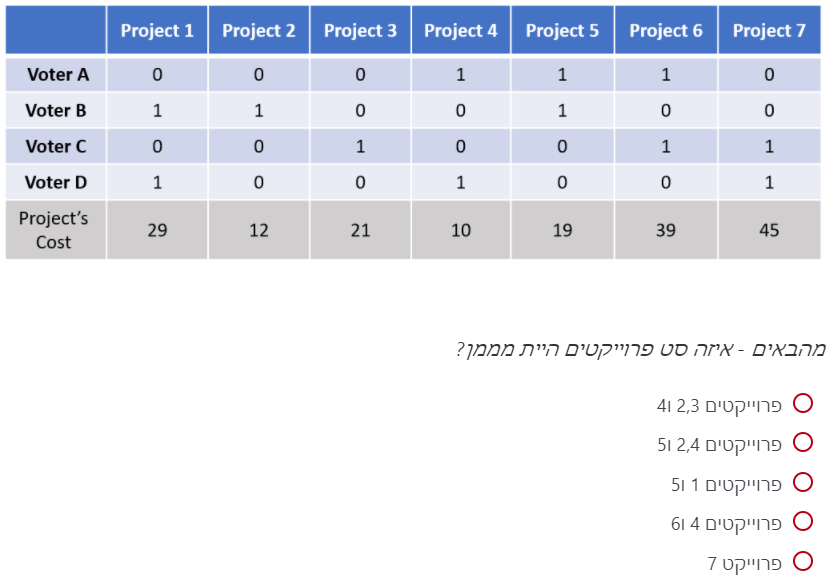}
\caption{Experiment 2-\textsc{SUM}: Scenario 4 out of 5}
\label{figure:twofour}
\end{figure}

\begin{figure}[t]
\centering
\includegraphics[width=0.85\textwidth]{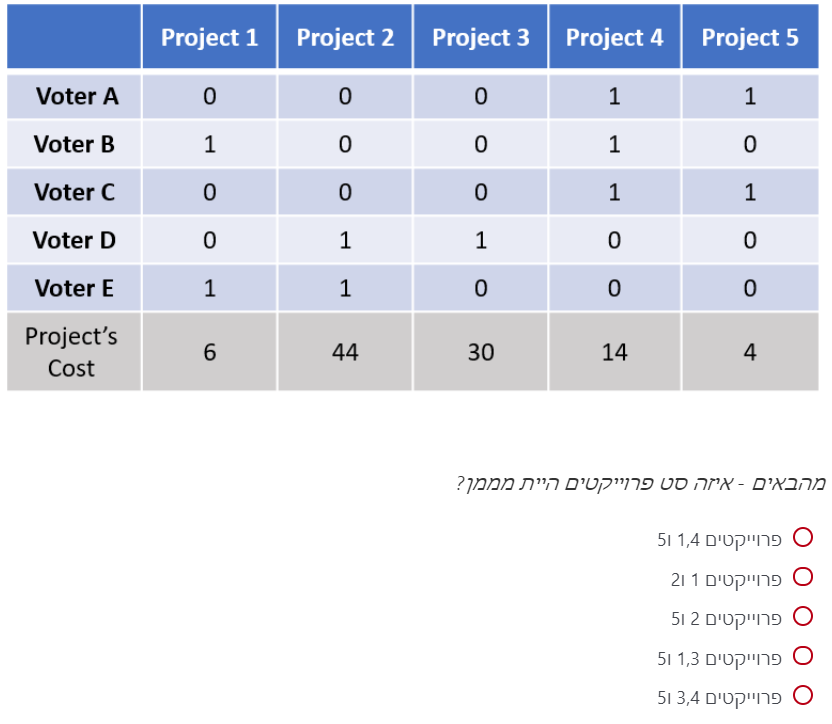}
\caption{Experiment 2-\textsc{SUM}: Scenario 5 out of 5}
\label{figure:twofive}
\end{figure}

The five instances used in Experiment 2-\textsc{SUM} (see Section \ref{section:experiment2}) appear in Figures~\ref{figure:twoone},~\ref{figure:twotwo},~\ref{figure:twothree},~\ref{figure:twofour}, and~\ref{figure:twofive}.

\begin{figure}[t]
\centering
\includegraphics[width=0.85\textwidth]{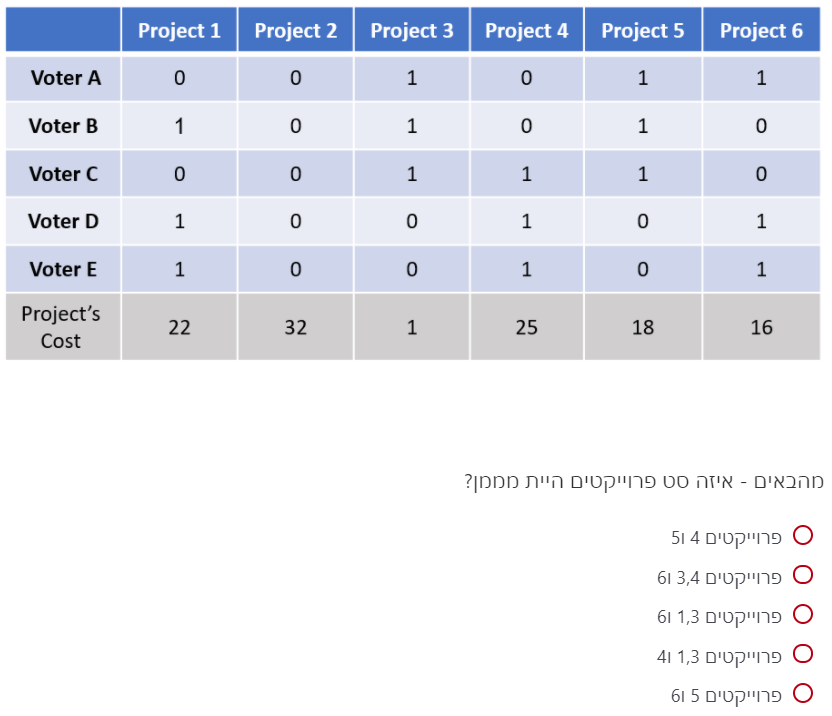}
\caption{Experiment 2-\textsc{NASH}: Scenario 1 out of 5}
\label{figure:threeone}
\end{figure}

\begin{figure}[t]
\centering
\includegraphics[width=0.85\textwidth]{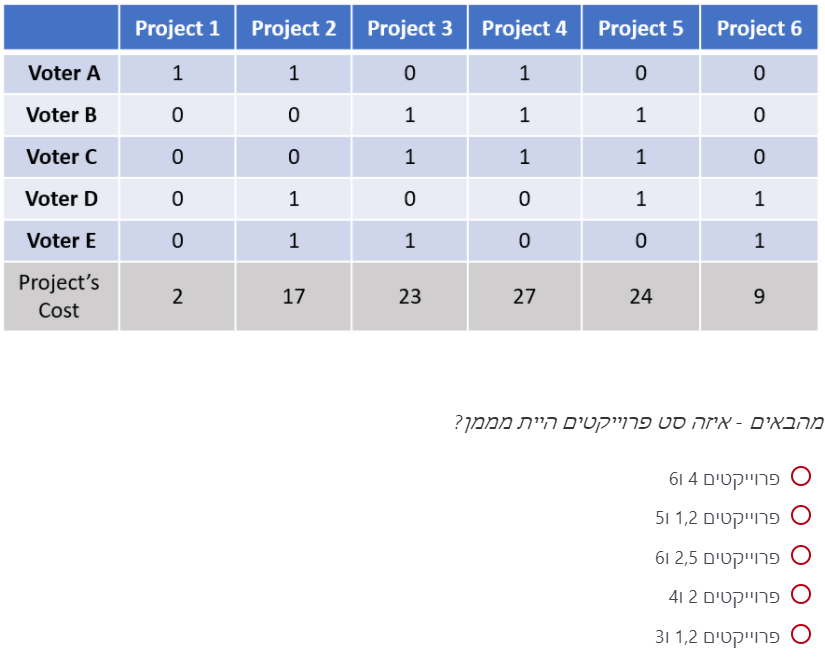}
\caption{Experiment 2-\textsc{NASH}: Scenario 2 out of 5}
\label{figure:threetwo}
\end{figure}

\begin{figure}[t]
\centering
\includegraphics[width=0.85\textwidth]{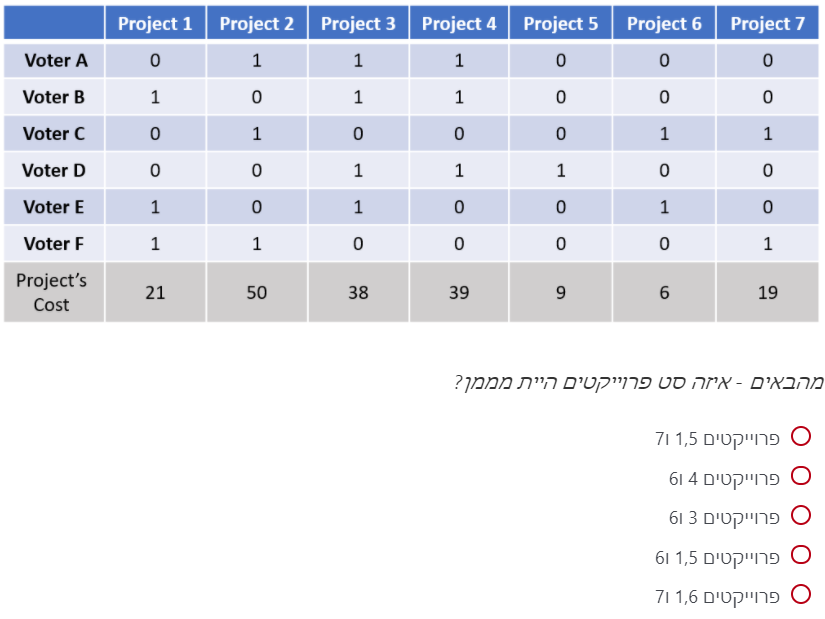}
\caption{Experiment 2-\textsc{NASH}: Scenario 3 out of 5}
\label{figure:threethree}
\end{figure}

\begin{figure}[t]
\centering
\includegraphics[width=0.85\textwidth]{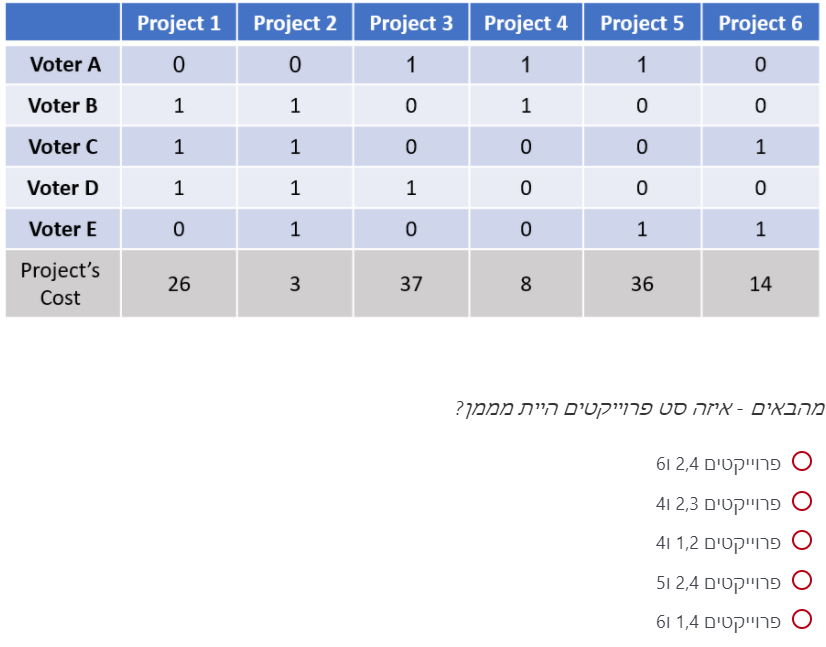}
\caption{Experiment 2-\textsc{NASH}: Scenario 4 out of 5}
\label{figure:threefour}
\end{figure}

\begin{figure}[t]
\centering
\includegraphics[width=0.85\textwidth]{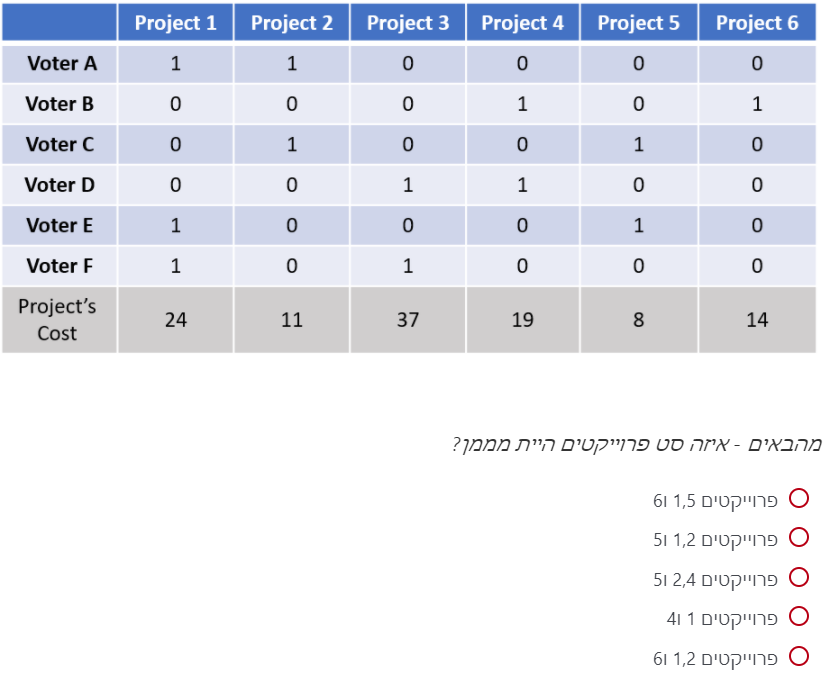}
\caption{Experiment 2-\textsc{NASH}: Scenario 5 out of 5}
\label{figure:threefive}
\end{figure}

The five instances used in Experiment 2-\textsc{NASH} (see Section \ref{section:experiment2}) appear in Figures~\ref{figure:threeone},~\ref{figure:threetwo},~\ref{figure:threethree},~\ref{figure:threefour}, and~\ref{figure:threefive}.

\end{document}